\documentclass[twocolumn,prl]{revtex4-1}

\usepackage{graphicx}
\usepackage{amsmath,amsthm}
\usepackage{amssymb}
\usepackage{layout}
\usepackage{bm}
\usepackage{bbold}
\usepackage{eufrak}
\usepackage{braket}
\usepackage{color}

\newtheorem{theorem}{Theorem}

\DeclareMathOperator{\dist}{dist}
\DeclareMathOperator{\diam}{diam}
\newcommand{\ic}{\ensuremath{\mathrm{i}}}
\renewcommand{\vec}[1]{\ensuremath{\bm{#1}}}
\newcommand{\ec}{\ensuremath{\mathrm{e}}}
\newcommand{\hilbert}{\ensuremath{\mathcal{H}}}

\date{\today}

\begin{document}

\title{Elementary excitations in gapped quantum spin systems}
\author{Jutho Haegeman$^1$, Spyridon Michalakis$^{2}$, Bruno Nachtergaele$^{3}$, Tobias J.~Osborne$^4$, Norbert Schuch$^{5}$, Frank Verstraete$^{1,6}$}
\affiliation{$^1$Department of Physics and Astronomy,  Ghent University, Ghent, Belgium\\
$^2$Institute for Quantum Information and Matter, Caltech, Pasadena, USA\\
$^2$Department of Mathematics, University of California, Davis, USA\\
$^4$Institute of Theoretical Physics, Leibniz Universit\"{a}t Hannover, Hannover, Germany\\
$^5$Institut f\"{u}r Quanteninformation, RWTH Aachen, Aachen, Germany\\
$^6$Vienna Center for Quantum Science, Universit\"at Wien, Wien, Austria}

\begin{abstract}
For quantum lattice systems with local interactions, the Lieb-Robinson bound serves as an alternative for the strict causality of relativistic systems and allows to prove many interesting results, in particular when the energy spectrum exhibits an energy gap. In this Letter, we show that for translation invariant systems, simultaneous eigenstates of energy and momentum with an eigenvalue that is separated from the rest of the spectrum in that momentum sector, can be arbitrarily well approximated by building a momentum superposition of a local operator acting on the ground state. The error satisfies an \emph{exponential} bound in the size of the support of the local operator, with a rate determined by the gap \emph{below and above} the targeted eigenvalue. We show this explicitly for the AKLT model and discuss generalizations and applications of our result.
\end{abstract}

\maketitle
Over 50 years ago, Zimmermann used the general principles of relativistic covariance and causality to show that the distinction between elementary excitations and bound states in relativistic quantum field theories is artificial and dependent on the formalism \cite{zimmermann}. While it is tempting to call excited states `elementary' when they can be connected to the one-particle excitations of the free theory by adiabatically switching off the interaction terms, in the fully interacting theory all discrete eigenstates of the energy-momentum operator $P_{\mu}$ are equivalent in that they can be created by acting on the vacuum with local operators satisfying the principle of microscopic causality. 

In non-relativistic quantum many body physics, such as quantum lattice systems, 
Zimmermann's result does not apply. Even for systems with translation invariance, there is no Lorentz boost to transform the 
energy spectrum and eigenstates in one momentum sector to a different momentum sector. Consequently, the complete energy 
momentum diagram and the set of excitations is often much more complex. 
In addition, in many strongly correlated systems, there is no obvious free theory that can serve as starting point to define the structure of elementary excitations. However, for systems with only local interactions in the Hamiltonian, the famous Lieb-Robinson (LR) bounds \cite{liebrobinson} can replace strict causality in \textit{e.g.} Fredenhagen's proof of the exponential clustering theorem \cite{fredenhagen}, which then results in a proof for the exponential decay of correlations for the ground state of any local gapped Hamiltonian \cite{clustering}. The LR bounds were also used in proving the area law for entanglement entropy of one-dimensional systems \cite{arealaw} and recent extensions thereof \cite{arealaw2}.

\begin{figure}
\includegraphics[width=\columnwidth]{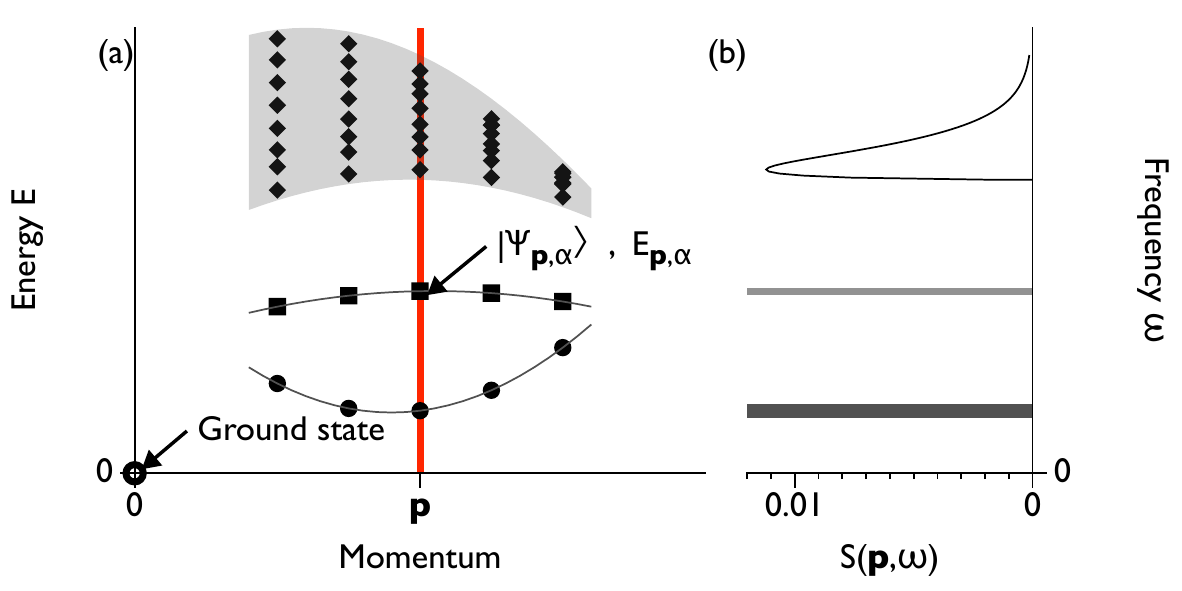}
\caption{Sketch of the energy momentum diagram. The points illustrate the eigenvalues of the Hamiltonian for a finite periodic system, whereas the lines and band illustrate the distribution of eigenvalues in the thermodynamic limit.}
\label{fig:sketch}
\end{figure}

In this Letter we extend the list of applications of the LR bounds by proving an analogue of Zimmermann's result for quantum lattice systems. For a translation invariant Hamiltonian, we show that an excitation energy that is separated by a gap from the rest of the spectrum within a given momentum sector corresponds to an eigenvector that can be approximated by the momentum superposition of a local operator acting on the ground state, with an error that is exponentially small in the size of the support of the operator. The picture to have in mind is sketched in FIG.~\ref{fig:sketch}. We target an energy eigenstate $\ket{\Psi_{\vec{p},\alpha}}$ with momentum $\vec{p}$ and an energy $E_{\vec{p},\alpha}$ that is separated from the other energy eigenvalues at momentum $\vec{p}$ by a gap that does not vanish in the thermodynamic limit. Because of the lack of Lorentz transformations, we cannot boost this state to any other momentum and have to treat each momentum sector independently. In particular, the resulting operator that is used to build an approximation of $\ket{\Psi_{\vec{p},\alpha}}$ can itself depend on the momentum $\vec{p}$. We also discuss possible extensions and applications of our result and provide a numerical illustration for the AKLT chain.

Let us consider a $d$-dimensional lattice $\Lambda$, which for simplicity we assume to be finite, cubic and periodic, \textit{i.e.}  $\Lambda\equiv \mathbb{Z}_{N}^{d}$. Arbitrary lattice sites are denoted by $\vec{x},\vec{y},\ldots\in\Lambda$. Sets of sites are denoted by $X,Y,\ldots$; the cardinality of a set $X$ is denoted as $\lvert X \rvert$. Let $\lVert \vec{x}\rVert$ denote the distance of a lattice point $\vec{x}\in\Lambda$ to the origin $\vec{o}\in\Lambda$, with $\lVert \cdot \rVert$ a suitable distance measure on $\mathbb{Z}_{N}^{d}$. The distance between two lattice points is $\dist(\vec{x},\vec{y})=\lVert \vec{y}-\vec{x}\rVert$; the distance between two sets $X$, $Y$ is defined as $\dist(X,Y)=\min_{x\in X,y\in Y} \dist(x,y)$; the diameter of a set $X$ is defined as $\diam(X)=\max_{x,y\in X}\dist(x,y)$. The Fourier transform of a lattice function $f:\Lambda\to\mathbb{C}$ is defined as
\begin{displaymath}
F(\vec{p})=\frac{1}{\sqrt{\lvert\Lambda\rvert}}\sum_{\vec{x}\in\Lambda} \ec^{-\ic \vec{p}\cdot \vec{x}} f(\vec{x})
\end{displaymath}
for any momentum vector $\vec{p}\in\mathcal{B}$, where the Brillouin zone $\mathcal{B}$ of a cubic lattice is given as $\mathcal{B}=\frac{2\pi}{N}\Lambda$. The inverse transformation is given by
\begin{displaymath}
f(x)=\frac{1}{\sqrt{\lvert\Lambda\rvert}}\sum_{\vec{p}\in\mathcal{B}} F(\vec{p})\, \ec^{\ic\vec{p} \cdot \vec{x}}.
\end{displaymath}

To all lattice sites $\vec{x}\in\Lambda$ we associate identical finite-dimensional Hilbert spaces $\hilbert_{\vec{x}}$; the Hilbert space of the whole system is $\hilbert_{\Lambda}=\bigotimes_{\vec{x}\in\Lambda} \hilbert_{\vec{x}}$. For all $\vec{x}\in\Lambda$, we define the space translation operator $T_{\vec{x}}$. The Hamiltonian of our system is given by $H_{\Lambda}=\sum_{X\subset \Lambda} H_X$, where the terms $H_X$ are supported on $X\subset\Lambda$. $H_{\Lambda}$ is assumed to be translation invariant: $\forall \vec{x} \in\Lambda: [T_{\vec{x}}, H_{\Lambda}]=0$. This is guaranteed by $H_{X+\vec{x}}=T_{\vec{x}} H_{X} T_{\vec{x}}^{\dagger}$. Furthermore, we assume that there exist constants $\mu, s>0$ for which
\begin{displaymath}
\sum_{X\ni \vec{x}} \lVert H_X\rVert \lvert X\rvert \exp[\mu \diam(X)]\leq s< \infty.
\end{displaymath}
This allows to use the LR bounds \cite{liebrobinson,lecturenotes}
\begin{multline}
\lVert [\ec^{\ic H_{\Lambda} t} A_X \ec^{-\ic H_{\Lambda} t},B_Y]\rVert \leq 2\lVert A_X\rVert \lVert B_Y \rVert\\
\times \lvert X\rvert \exp\left[-\mu \dist(X,Y)\right] \left[\exp(2 s \lvert t\rvert)-1\right]
\label{eq:lrbound}
\end{multline}
for two operators $A_X$ and $B_Y$ supported on disjoint sets $X$ and $Y$. For simplicity, we first assume that $H_{\Lambda}$ has a unique ground state $\ket{\Psi_0}\in\hilbert_{\Lambda}$ that is necessarily translation-invariant (\textit{i.e.} momentum $\vec{p}=\vec{0}$). The energy scale is chosen such that the ground state energy is zero. We discuss the case of ground state degeneracy further on. Important is the existence of a spectral gap $\Delta E$ that separates the ground state energy from the rest of the energy spectrum and does not vanish in the thermodynamic limit. All other eigenstates of $H$ are denoted as $\ket{\Psi_{\vec{p},\alpha}}$ with energy $E_{\vec{p},\alpha}\geq \Delta E$, where $\vec{p}\in \mathcal{B}$ labels the momentum sector and the index $\alpha$ labels all eigenstates within a given momentum sector. They are assumed to be normalized such that $\braket{\Psi_{\vec{p},\alpha}|\Psi_{\vec{p}',\alpha'}}=\delta_{\vec{p},\vec{p}'}\delta_{\alpha,\alpha'}$.

As a last preliminary, we introduce two more definitions before stating the main theorem. The Fourier transform of an operator $O$, denoted as $O(\vec{p})$, is defined by
\begin{displaymath}
O(\vec{p})=\frac{1}{\sqrt{\lvert \Lambda\rvert}}\sum_{\vec{x}\in \Lambda} \ec^{\ic \vec{p} \cdot \vec{x}} T_{\vec{x}} O T_{\vec{x}}^{\dagger}
\end{displaymath}
For every bounded operator $O$ with support in the compact set $X$ and with zero ground state expectation value ($\braket{\Psi_0|O|\Psi_0}=0$), we can define a momentum eigenstate $\ket{\Phi_p[O]}$ with momentum $\vec{p}$ by
\begin{equation}\label{defn:momentum_eigenstate}
\ket{\Phi_{\vec{p}}[O]}=O(\vec{p})\ket{\Psi_0}=\frac{1}{\sqrt{\lvert \Lambda\rvert}}\sum_{\vec{x}\in \Lambda} \ec^{\ic \vec{p} \cdot \vec{x}} T_{\vec{x}} O\ket{\Psi_0}.
\end{equation}
The exponential decay of connected correlation functions set by the spectral gap $\Delta E$ makes it straightforward to show that the norm of $\ket{\Phi_{\vec{p}}[O]}$ remains finite in the thermodynamic limit. In particular, for an operator $O_X$ supported on $X \subset \Lambda$, the following bound follows from~\cite{clustering}:
\begin{equation}\label{bound:exp_cluster}
\lVert\ket{\Phi_{\vec{p}}[O_X]}\rVert \le \sqrt{\diam(X)+ |X|/\delta}\, \lVert O_X \rVert,
\end{equation}
with $\delta \sim \Delta E$ independent of the lattice size.
In the formulation of our main result below we use the notation $B_\ell(X)$
for the set of sites $x$ for which $\dist(x,X)\leq \ell$. 

\begin{theorem}
Let $\ket{\Psi_{p,\alpha}}$ be a normalized momentum $p$ eigenstate of $H$ with a non-degenerate eigenvalue $E_{p,\alpha}$, such that in the thermodynamic limit $\lvert \Lambda\rvert\to \infty$:
\begin{displaymath}
\forall \beta\neq \alpha: \lvert E_{p,\alpha}-E_{p,\beta}\rvert \geq \delta E
\end{displaymath}
In addition, we assume there exists an operator $O$ with $\braket{\Psi_0|O|\Psi_0}=0$ and support in $X$, such that its Fourier transform $O(p)$ has a non-zero \emph{spectral weight} $\lvert \braket{\Psi_{p,\alpha}|O(p)|\Psi_0}\rvert \geq f \lVert O\rVert$, with $f>0$ independent of $\lvert \Lambda\rvert$.

We can then define a new operator $O^{(\ell)}$ with support in $B_\ell(X)$, such that 
$\ket{\Phi_{p}[O^{(\ell)}]}$ defined in~\eqref{defn:momentum_eigenstate}, satisfies:
\begin{equation}
F=\frac{\lvert\braket{\Psi_{p,\alpha}|\Phi_{p}[O^{(\ell)}]}\rvert}{\lVert\ket{\Phi_{p}[O^{(\ell)}]}\rVert}\geq 1-p_X(\ell) \exp\left[-\frac{\delta E}{2v_{\mathrm{LR}}}\ell\right]\label{eq:fidelity}
\end{equation}
for $\ell$ sufficiently large, a polynomial $p_X(\ell) \sim D_X(\ell)\, f^{-1}$ with $D_X(\ell)$ defined in~\eqref{defn:D_X}, and $v_{\mathrm{LR}} = (\delta E/2 +2s)/\mu$.
\end{theorem}

Hence, we can approximate the excited state by acting with the momentum superposition of a localized operator $O^{(\ell)}$ with an error that is exponentially small in the linear size of the support of the operator. Before continuing to the proof, some comments are in order. We assume that a local operator $O$ can be found for which $\lvert\braket{\Psi_{p,\alpha}|O(p)|\Psi_0}\rvert\geq f \lVert O\rVert$ is nonzero. This factor appears as the strength of the isolated pole $\omega=E_{\alpha}$ in the (Fourier-transformed) dynamic correlation function
\begin{displaymath}
D(\vec{p},\omega)=\braket{\Psi_0|O(\vec{p})^{\dagger} \frac{1}{\omega-H_{\Lambda}+\ic \epsilon}O(\vec{p})|\Psi_0}.
\end{displaymath}
These poles appear as $\delta$-singularities in the spectral function $S(\vec{p},\omega)\sim\mathrm{Im} D(\vec{p},\omega)$. The non-triviality of our result is in the fact that $f$ can be an arbitrarily small fraction and most of the spectral weight can be distributed at other isolated poles or at a continuum in the thermodynamic limit, as sketched in FIG.~\ref{fig:sketch}(b). The rate of exponential convergence of $F\to 1$ as a function of $\ell$ does not depend on the magnitude of $f$. The importance of connecting excitations of the system to the spectral function of local operators is that this is the main observable from which information about excitations can be obtained in experiments. Note that we are only considering isolated singularities of $S(\vec{p},\omega)$, which correspond to eigenstates of the Hamiltonian. We are not considering quasi-particles in the sense of strong resonances in the continuum distribution of the spectral function,
which are superpositions of many eigenstates with slightly different energies and thus only have a finite lifetime \cite{quasiparticles}. 

\begin{proof}
We start by applying an energy filter to define $O_1$:
\begin{equation}
O_1=\frac{1}{\sqrt{2\pi q}}\int_{-\infty}^{+\infty} \ec^{-\ic H t} O \ec^{+\ic H t} \ec^{\ic E_{\alpha,p} t}\ec^{-\frac{t^2}{2q}}\,\mathrm{d} t
\end{equation}
so that the state $\ket{\Phi_{\vec{p}}[O_1]}$ satisfies:
\begin{displaymath}
\lvert \braket{\Psi_{\vec{p},\alpha}|\Phi_{\vec{p}}[O_1]}\rvert =
\lvert \braket{\Psi_{\vec{p},\alpha}|\Phi_{\vec{p}}[O]}\rvert \geq f \lVert O\rVert
\end{displaymath}
and for any $\beta\neq \alpha$:
\begin{displaymath}
\lvert \braket{\Psi_{\vec{p},\beta}|\Phi_{\vec{p}}[O_1]}\rvert \leq \ec^{-\frac{q \delta E^2}{2}} \lvert \braket{\Psi_{\vec{p},\beta}|\Phi_{\vec{p}}[O]}\rvert.
\end{displaymath}
By restricting the time-integration in the energy-filtering, we obtain a new operator $O_2$
\begin{equation}
O_2=\frac{1}{\sqrt{2\pi q}}\int_{-T}^{+T} \ec^{-\ic H t} O \ec^{+\ic H t} \ec^{\ic E_{\alpha,p} t}\ec^{-\frac{t^2}{2q}}\,\mathrm{d} t.
\end{equation}
Using the triangle inequality we obtain the bound
\begin{align*}
\lvert \braket{\Psi_{\vec{p},\alpha}|\Phi_{\vec{p}}[O_2]}\rvert \geq
\lvert \braket{\Psi_{\vec{p},\alpha}|\Phi_{\vec{p}}[O_1]}\rvert-\lvert \braket{\Psi_{\vec{p},\alpha}|\Phi_p[O_2-O_1]}\rvert\\
\geq \left(1-c\,\ec^{-\frac{T^2}{2q}}\right)
\lvert \braket{\Psi_{\vec{p},\alpha}|\Phi_{\vec{p}}[O]}\rvert\geq \left(1-c\,\ec^{-\frac{T^2}{2q}}\right) f \lVert O \rVert
\end{align*}
and for $\beta\neq \alpha$:
\begin{align*}
\lvert \braket{\Psi_{\vec{p},\beta}|\Phi_{\vec{p}}[O_2]}\rvert &\leq
\lvert \braket{\Psi_{\vec{p},\beta}|\Phi_{\vec{p}}[O_1]}\rvert+\lvert \braket{\Psi_{\vec{p},\beta}|\Phi_{\vec{p}}[O_2-O_1]}\rvert\nonumber\\
&\leq \left(\ec^{-\frac{q\delta E^2}{2}}+c\,\ec^{-\frac{T^2}{2q}}\right) \lvert \braket{\Psi_{p,\beta}|\Phi_p[O]}\rvert, 
\end{align*}
with $c$ such that $(\pi)^{-1/2}\int_{\lvert t\rvert >T}\ec^{-t^2}\,\mathrm{d}t\leq c\,\ec^{-T^2}$. Finally, we replace $O_2$ by its localized version $O^{(\ell)}$
\begin{equation}
O^{(\ell)}={\rm Tr}_{\hilbert_{\Lambda\setminus B_\ell(X)}} O_2,
\end{equation}
by taking the normalized partial trace over the spins in $\Lambda\setminus B_{\ell}(X)$. From the LR bounds in Eq.~\eqref{eq:lrbound}, we obtain:

\begin{displaymath}
\lVert O^{(\ell)}-O_2\rVert\leq \frac{2\lvert X\rvert}{s \sqrt{2\pi q}} \lVert O\rVert \exp(2s T-\mu \ell).
\end{displaymath}  
Moreover, we have $O_2 - O^{(\ell)} = \sum_{n \ge \ell} (O^{(\ell+1)}-O^{(\ell)})$, with $\|O^{(\ell+1)}-O^{(\ell)}\| \le \|O^{(\ell+1)}-O_2\| + \|O_2-O^{(\ell)}\| \le \frac{4\lvert X\rvert}{s \sqrt{2\pi q}} \lVert O\rVert \exp(2s T-\mu \ell)$.
Together with the bound~\eqref{bound:exp_cluster}, from the telescoping sum we get: 
\begin{eqnarray*}
&\lVert \ket{\Phi_{\vec{p}}[O^{2}-O^{(\ell)}]}\rVert \leq \sum_{n\geq \ell} \lVert \ket{\Phi_{\vec{p}}[O^{(n+1)}-O^{(n)}]}\rVert \leq \\ 
&\sum_{n\geq \ell} C\big(B_{n+1}(X)\big) \lVert O^{(n+1)}-O^{(n)}\rVert \leq  D_X(\ell) \lVert O\rVert e^{2s T-\mu \ell},
\end{eqnarray*}
with $C(Y) \sim \sqrt{\diam(Y)+ |Y|/\delta}, \, \delta \sim \Delta E$ and
\begin{equation}\label{defn:D_X}
D_X(\ell) \sim \lvert X\rvert \, C\big(B_{\ell+1}(X)\big)/(s\, \mu\, \sqrt{2\pi q}).
\end{equation}
Using the above bound:
{\small
\begin{align*}
\lvert& \braket{\Psi_{\vec{p},\alpha}|\Phi_{\vec{p}}[O^{(\ell)}]}\rvert \geq
\lvert \braket{\Psi_{\vec{p},\alpha}|\Phi_p[O_2]}\rvert-\lvert \braket{\Psi_{\vec{p},\alpha}|\Phi_{\vec{p}}[O^{(\ell)}-O_2]}\rvert\nonumber\\
&\geq \left(1- c\,\ec^{-\frac{T^2}{2q}}-(D_X(\ell)/f)  e^{2s T-\mu \ell}\right)f \lVert O\rVert.
\end{align*}}\noindent
Define the seminorm $\lVert\ket{\Phi}\rVert' =(\sum_{\beta\neq\alpha} \lvert\braket{\Psi_{p,\beta}|\Phi}\rvert^2)^{1/2}$, which satisfies $\lVert\ket{\Phi}\rVert'\leq \lVert \ket{\Phi}\rVert$. Then
\begin{align*}
\lVert&\ket{\Phi_{\vec{p}}[O^{(\ell)}]}\rVert'\leq
\lVert \ket{\Phi_{\vec{p}}[O_2]}\rVert'+\lVert\ket{\Phi_{\vec{p}}[O^{(\ell)}-O_2]}\rVert'\nonumber\\
&\leq \left(\ec^{-\frac{q\delta E^2}{2}}+c\,\ec^{-\frac{T^2}{2q}}\right) \lVert\ket{\Phi_p[O]}\rVert
+\lVert\ket{\Phi_{\vec{p}}[O_2-O^{(\ell)}]}\rVert\nonumber\\
&\leq \left(\ec^{-\frac{q\delta E^2}{2}}+c \ec^{-\frac{T^2}{2q}}+D_X(\ell) e^{2s T-\mu \ell}\right) \lVert O\rVert.
\end{align*}

Since we are trying to construct a lower bound for the fidelity $F$ from Eq.~\eqref{eq:fidelity}, we can upper bound the denominator by
$\lVert\ket{\Phi_{p}[O^{(\ell)}]}\rVert\leq \lvert\braket{\Psi_{p,\alpha}|\Phi_{p}[O^{(\ell)}]}\rvert+\lVert \ket{\Phi_p[O^{(\ell)}]}\rVert'$ so that:
\begin{displaymath}
F\geq \frac{1}{1+\frac{\lVert\ket{\Phi_p[O^{(\ell)}]}\rVert'
}{\lvert\braket{\Psi_{p,\alpha}|\Phi_{p}[O^{(\ell)}]}\rvert}}\geq 1-\frac{\lVert\ket{\Phi_p[O^{(\ell)}]}\rVert'
}{\lvert\braket{\Psi_{p,\alpha}|\Phi_{p}[O^{(\ell)}]}\rvert}
\end{displaymath}
Using an upper bound for the numerator of the second term, and a lower bound for its denominator, we get 
\begin{displaymath}
F\geq 1-\frac{1}{f}\frac{\ec^{-\frac{q\delta E^2}{2}}+c \ec^{-\frac{T^2}{2q}}+D_X(\ell) \,e^{2s T-\mu \ell}}{1- c\,\ec^{-\frac{T^2}{2q}}-(D_X(\ell)/f)\,  e^{2s T-\mu \ell}}.
\end{displaymath}
We now set $T=\ell/v_{\text{LR}}$, $q=T/\delta E$ and $v_{\text{LR}}$ such that $\mu-2s/v_{\text{LR}} = \delta E/2v_{\text{LR}}$ in order to reproduce Eq.~\eqref{eq:fidelity}:
\begin{equation}
F\geq 1-\left(\frac{1+c +D_X(\ell)}{1- (c + D_X(\ell)/f)\,\ec^{-\frac{\delta E}{2v_{\text{LR}}}\ell}} \right)\frac{1}{f}\,\ec^{-\frac{\delta E}{2v_{\text{LR}}}\ell},
\end{equation}
with $\ell \ge \ell_0$, where $(c + D_X(\ell)/f)\,\ec^{-\frac{\delta E}{2v_{\text{LR}}}\ell_0} \le 1/2$.
\end{proof}

It is easy to generalize the result above in the case where the energy eigenvalue $E_{\vec{p},\alpha}$ is (nearly) 
degenerate, or overlaps with other eigenvalues, but where we can isolate it by restricting to a sector with specific 
quantum numbers corresponding to additional symmetries of the Hamiltonian. For example, to approximate an 
isolated spin-$J$ multiplet of excitations of an $\mathsf{SU}(2)$ symmetric Hamiltonian, we can use a set of
operators $O_{m}$ ($m=-J,\ldots,+J$) that also transform as the spin $J$ irreducible representation under 
$\mathsf{SU}(2)$.

The case of a degenerate ground state is more subtle and requires an approach in terms of operator algebras and their 
representations. Similar developments in the context of the Haag-Kastler framework of local quantum theory \cite{lqt} 
do reproduce the locality of excitations \cite{lqtlocal}, but it was also found that some excitations require the introduction 
of non-local fields in string-like regions \cite{lqtstring}. Indeed, for every ground state of the system, the GNS construction
provides a Hilbert space in which we can apply the same formalism, with possible modification to accommodate for the 
fact that the ground state might not be translation invariant. Note that the construction above survives the thermodynamic
limit, the only complication being that momentum eigenstates can no longer be normalized to $1$ and should satisfy a 
$\delta$-normalization instead. However, there might be additional representations, that do not correspond to a single 
ground state and which define unitarily inequivalent Hilbert spaces, known as superselection sectors, in which to look 
for excitations. These stringlike excitations appear as kinks in systems with symmetry breaking \cite{kinks}, as electric 
charges in gauge theories \cite{gauge} or as anyonic excitations in systems with topological order
\cite{braidgroup,latticeanyons}.

\begin{figure}
\includegraphics[width=\columnwidth]{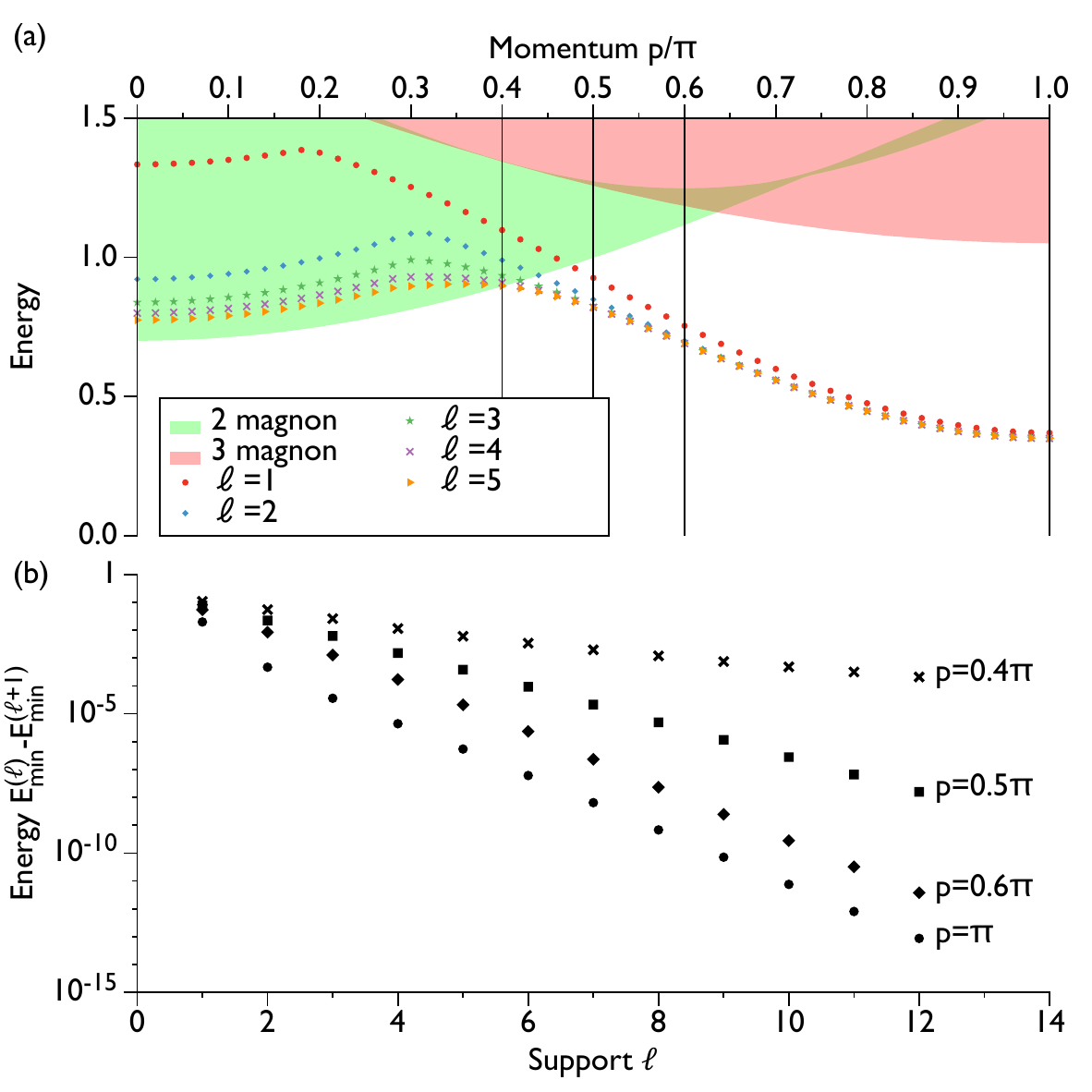}
\caption{Lowest variational excitation energies $E^{(\ell)}_{\text{min}}$ obtained with the ansatz from Eq.~\eqref{eq:akltansatz}. Subpanel (a) shows the $E^{(\ell)}_{\text{min}}$ as function of the momentum $p$ for $\ell=1,\ldots,5$, as well as approximate position of the 2-magnon and 3-magnon continuum based on the numerical results for the one-magnon dispersion relation with $\ell=5$. Subpanel (b) illustrates the exponential convergence of $E_{\min}^{(\ell)}$ by plotting $E^{(\ell)}_{\min}-E^{(\ell+1)}_{\min}$ for different values of $p$ [as indicated by vertical lines in subpanel (a)].}
\label{fig:aklt}
\end{figure} 

The main application of our result is that it validates constructions such as the Feynman-Bijl ansatz \cite{feynmanbijl} or the single mode approximation \cite{sma} as variational ansatz for excitations. Given the exact ground state $\ket{\Psi_0}$, we can try to determine the spectrum of isolated excitations by building a linear space of states $\ket{\Phi_{\vec{p}}[O]}$, where the variational parameters are encoded in the operator $O$ with support on a compact set $X$. That this even works with an {\em approximate} ground state $\ket{\tilde{\Psi}_0}$ was illustrated within the framework of matrix product states (MPS) \cite{mps,smamps}. For the particular example of the AKLT model \cite{aklt}, which has an exact MPS ground state with 
bond dimension $D=2$, we illustrate the exponential convergence of the excitation energy of a trial state of the form
\begin{equation}
\sum_{n\in\mathbb{Z}}\sum_{\vec{s}} \ec^{\ic p n}T_{n} \vec{v}_{\mathrm{L}}^{\dagger} \cdots A^{s_{0}} B^{s_{1}s_{2}\ldots s_{\ell}} A^{s_{\ell+1}}\cdots \bm{v}_{\mathrm{R}}\ket{\bm{s}}\label{eq:akltansatz}
\end{equation}
where $\vec{s} = (s_0,s_1,\ldots)$ with $s_k=\{-1,0,+1\}$, the matrices $A^{s_k}$ encode the ground state, the tensor $B$ acts on a block of $\ell$ sites and contains the variational parameters, $\ket{\bm{s}}=\cdots \otimes\ket{s_{n}}\otimes\ket{s_{n+1}}\otimes\cdots$ is the direct product basis and $\bm{v}_{\mathrm{L,R}}$ are boundary vectors which disappear in the thermodynamic limit. The variational space spanned by this ansatz is equivalent to the set of states $\ket{\Phi_{p}[O]}$ where $O$ acts on $\ell$ sites. FIG.~\ref{fig:aklt}(a) shows the lowest 3-fold degenerate excitation energy across momentum space, whereas FIG.~\ref{fig:aklt}(b) shows the convergence of this energy as a function of $\ell$. For the selected momenta $p>0.4\pi$, the variational energy converges exponentially fast as a function of the block size $\ell$. Momentum $p=0.4\pi$ is a borderline case, as it is hard to predict from the numerics whether the gap between the single-magnon dispersion curve and the multi-magnon continuum is still open. Indeed, the absence of Lorentz boost symmetries allows excitations to only exist in certain subdomains of momentum space.

If the ground state of some Hamiltonian $H_{\Lambda}$ is an exact MPS, one can use the injectivity property of MPS \cite{mps} to show that minimizing the energy with respect to a local operator $O$ acting on $\ell$ sites is equivalent to finding energy eigenstates of the same Hamiltonian on a lattice of $\ell$ sites with 2 added boundary sites and corresponding boundary terms. One can then show the existence of local operators $O$ with the required properties by bounding the energy of the special boundary terms \footnote{S.~Michalakis \textit{et. al.}, in preparation.}. Note that Theorem~1 also allows us to conclude that ---in the thermodynamic limit--- the entanglement entropy of isolated excitations equals that of the ground state plus $\log(2)$, corresponding to the excitation being left or right to the cut. The contribution of the terms
spanning the cut vanishes in the thermodynamic limit, in agreement with Ref.~\onlinecite{pizorn}. 

In conclusion, we have shown that excited states of translation invariant lattice Hamiltonians for which the energy eigenvalue is isolated within a given momentum sector, and which can be detected in the spectral function of local operators, can be arbitrarily well approximated by the momentum superposition of a local operator acting on the ground state. By identifying these `elementary' excitations with single particle states, we will show in a later publication that ---in the thermodynamic limit--- we can then build a Hilbert space of multi-particle excitations starting from the fully interacting vacuum, and in particular, that we can formulate the scattering problem for such excitations \footnote{J.~Haegeman \textit{et.\ al.}, in preparation.}.

\begin{acknowledgments}
Discussions with Karel Van Acoleyen and Henri Verschelde are greatly acknowledged. We acknowledge funding provided by the Institute for Quantum Information and Matter, an NSF Physics Frontiers Center with support of the Gordon and Betty Moore Foundation through Grant \#GBMF1250 and by the AFOSR Grant \#FA8750-12-2-0308 (SM), the National Science Foundation under grant DMS-1009502 (BN), the Alexander von Humboldt foundation (NS), an Odysseus grant from the FWO Flanders (FV,JH), the FWF grants FoQuS and Vicom (FV), the ERC grants QUERG (FV) and QFTCMPS (TJO) and by the cluster of excellence EXC 201 Quantum Engineering and Space-Time Research (TJO).
\end{acknowledgments}

\end{document}